# SECURED GREEDY PERIMETER STATELESS ROUTING FOR WIRELESS SENSOR NETWORKS

P. Samundiswary[1], D.Sathian[1] and P. Dananjayan[2]

[1]Department of Electronics and Communication Engineering and Computer Science and Engineering
Sri Manakula Vinayagar Engineering College
Pondicherry-605107, India
[2]Department of Electronics and Communication Engineering
Pondicherry Engineering College
Pondicherry-605014, India
pdananjayan@rediffmail.com

## ABSTRACT

*Wireless sensor networks are collections of large number of sensor nodes. The sensor nodes are featured with limited energy, computation and transmission power. Each node in the network coordinates with every other node in forwarding their packets to reach the destination. Since these nodes operate in a physically insecure environment; they are vulnerable to different types of attacks such as selective forwarding and sinkhole. These attacks can inject malicious packets by compromising the node. Geographical routing protocols of wireless sensor networks have been developed without considering the security aspects against these attacks. In this paper, a secure routing protocol named secured greedy perimeter stateless routing protocol (S-GPSR) is proposed for mobile sensor networks by incorporating trust based mechanism in the existing greedy perimeter stateless routing protocol (GPSR). Simulation results prove that S-GPSR outperforms the GPSR by reducing the overhead and improving the delivery ratio of the networks.*

## 1. INTRODUCTION

Recently wireless sensor networks have drawn a lot of attention due to broad applications in military and civilian operations. Sensor nodes in the network are characterized by severely constrained, energy resources and communicational capabilities. Due to small size and unattention of the deployed nodes, attackers can easily capture and rework them as malicious nodes. Karlof and Wagner [1] also have revealed that routing protocols of sensor networks are insecure and highly vulnerable to malicious nodes. It can either join the network externally or may originate internally by compromising an existing benevolent node [2]. The attacks launched by internally generated compromised nodes are the most dangerous type of attacks. These compromised nodes can also carry out both passive and active attacks against the networks [3]. In passive attack a malicious node only eavesdrops upon the packet contents, while in active attacks it may imitate, drop or modify legitimate packets [4]. Sinkhole is one of the common type of active attack [5] in which a node, can deceitfully modify the routing packets. So, it may lure other sensor nodes to route all traffic through it. The impact of sinkhole is to launch further active attacks on the traffic, which is routed through it.

Due to limited capabilities of sensor nodes, providing security and privacy against these attacks is a challenging issue to sensor networks. In order to protect network against malicious attackers, numbers of routing protocols have been developed to improve network performance with the help of cryptographic techniques. Security mechanisms used in these routing protocols of sensor networks detect the compromised node and then revoke the cryptographic keys of the network. But, requirements of such secure routing protocols include configuration of the nodes with encryption keys and the creation of a centralized or distributed key repository to realize different security services in the network [6].

In addition, secure routing protocols utilising cryptographic methods also require excessive overheads. However, only few routing protocols such as secured dynamic source routing protocols (S-DSR) for wireless sensor networks address the security mechanism by using trust based model against various attacks [7]. S-DSR forwards the packets to successive nodes given in the source node route header by checking its trust levels only. S-DSR will not utilise geographic position of the neighbour node closest to the destination to forward the packet. Greedy perimeter stateless routing is one of the protocols which transmits packet by using the position of neighbour node with respect to the destination node. GPSR uses two methods such as greedy forwarding and perimeter forwarding mechanism to transmit data from source to destination. But GPSR is also exposed to various types of attackers. In this paper, secured greedy perimeter stateless routing (S-GPSR) is implemented by including trust mechanism and mobility model for nodes in greedy perimeter stateless routing to protect nodes from sinkhole attacks. This S-GPSR is simulated by using ns-2.32 for different coverage areas of 300m×300m and 500m×500m with 150 and 200 numbers of nodes considering mobile nodes in the network. The paper is organized as follows: Section 2 describes about the greedy perimeter stateless routing. Section 3 deals with the proposed secured greedy perimeter stateless routing of wireless sensor networks. Simulation results are discussed in Section 4 to obtain delivery ratio, delay and routing overhead of the proposed security scheme and conclusions are drawn in Section 5.

## 2. GREEDY PERIMETER STATELESS ROUTING

Routing in sensor networks is very challenging due to several characteristics that distinguish them from contemporary communication and wireless ad-hoc networks. First of all, it is not possible to build a global addressing scheme for the deployment of sheer number of sensor nodes. Therefore, traditional IP-based protocols cannot be applied to sensor networks. Second, in contrary to typical communication networks almost all applications of sensor networks require the flow of sensed data from multiple sources to a particular sink. Third, generated data traffic has significant redundancy in it since multiple sensors may generate same data within the vicinity of a phenomenon. Such redundancy needs to be exploited by routing protocols to improve energy and bandwidth utilization. Fourth, sensor nodes are tightly constrained in terms of transmission power, on-board energy, processing capacity and storage and thus require careful resource management.

Due to the above differences, many new algorithms have been proposed for the problem of routing data in sensor networks. These routing mechanisms have considered the characteristics of sensor nodes along with the application and architecture requirements. Almost all routing protocols can be classified as data-centric, hierarchical or location-based although there are few distinct ones based on network flow or QoS awareness. Data-centric protocols are query-based and depend on the naming of desired data, which helps in eliminating many redundant transmissions. Hierarchical protocols aim at clustering the nodes so that cluster heads can do some aggregation and reduction of data in order to save energy [8]. Location-based protocols utilize the position information to relay the data to desired regions rather than the whole network.

The Greedy Perimeter Stateless Routing is one of the commonly used location-based routing protocols for establishing and maintaining a sensor network. This protocol virtually operates in a stateless manner and has the ability for multi-path routing. In GPSR, it is assumed that all nodes recognise the geographical position of destination node with which communication is desired. This location information (i.e.) geographical position is also used to route traffic to its requisite destination from the source node through the shortest path. Each transmitted data packet from node contains the destination node's identification and its geographical position in the form of two four-byte float numbers. Each node also periodically transmits a beacon, to inform its adjacent nodes regarding its current geographical co-ordinates. The node positions are recorded, maintained and updated in a neighbourhood table by all nodes receiving the beacon. To reduce the overhead due to periodic beacons, the node positions are piggy-backed onto forwarded data packets.

GPSR supports two mechanisms for forwarding data packets: greedy forwarding and perimeter forwarding [9].

*i) Greedy Forwarding*

In the first mechanism, all data packets are forwarded to an adjacent neighbour that is geographically positioned closer to the intended destination. This mechanism is known as greedy forwarding. The forwarding is done on a packet to packet basis. Hence, minimal state information is required to be retained by all nodes. It makes protocol most suitable for resource starved devices. The greedy forwarding mechanism is shown in Figure1. However, this mechanism is susceptible to failure in situations where the distance between forwarding node and final destination is less than the distance between the forwarding node's adjacent neighbours and destination.

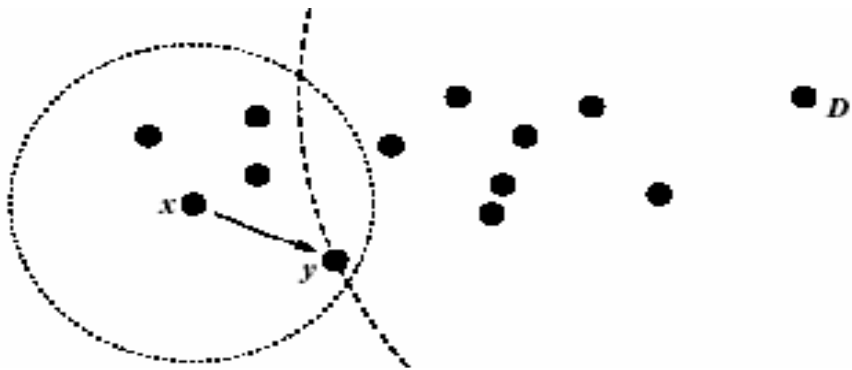

Figure 1. Greedy forwarding mechanism

*ii) Perimeter Forwarding*

To overcome routing problems in such scenarios, GPSR engages perimeter forwarding mode. In perimeter mode, the data packet is marked as being in perimeter mode along with the location where greedy forwarding failed. These perimeter mode packets are forwarded using simple planar graph traversal. Each node receiving a data packet marked as in perimeter mode uses the right-hand rule to forward packets to nodes, which are located counterclockwise to the line joining forwarding node and the destination. The perimeter forwarding mechanism is shown in Figure 2. Each node, while forwarding perimeter mode packets, compares its present distance to the destination from the point where greedy forwarding has failed. If the current distance is less, packet is routed through greedy forwarding repeatedly from that point onwards.

The protocol has been designed and developed based on the assumption that all nodes in the network would execute the protocol in a sincere manner. However, due to number of reasons including malice, incompetence and selfishness, nodes frequently deviate from defined standards leading to routing predicaments.

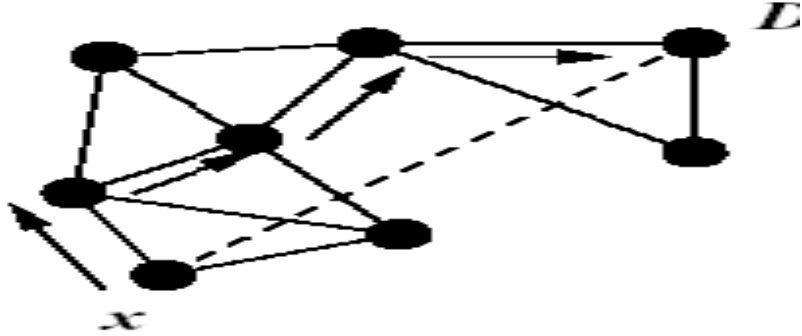

Figure 2. Perimeter forwarding mechanism

## 3. SECURED GREEDY PERIMETER STATELESS ROUTING PROTOCOL Of WIRELESS SENSOR NETWORKS

GPSR scans its neighbourhood table to retrieve the next hop which is optimal and leads to the destination, during packet transmission to a known host. As there may be more than one such hop available, GPSR selects an adjacent neighbour that has the least distance to a particular destination. In S-GPSR, the trust levels used in conjunction with the geographical distances are incorporated in the neighbourhood table to create the most trusted distance route rather than the default minimal distance. To compute direct trust in a node, an effort-return based trust model is used [10]. The accuracy and sincerity of immediate neighbouring nodes is ensured by observing their contribution to packet forwarding mechanism.

To implement the trust derivation mechanism, Trust Update Interval (TUI) of each forwarded packet is buffered in the node as (GPSR Agent::buffer packet). The TUI is a very critical component of such a trust model. It determines the time a node should wait before assigning a trust or distrust level to a node based upon the results of a particular event. After transmission, each node promiscuously listens for the neighbouring node to forward the packet. If neighbour forwards the packet in proper manner within the TUI, its corresponding trust level is incremented. However, if the neighbouring node modifies the packet in an unexpected manner or does not forward the packet at all, its trust level is decremented.

Every time a node transmits a data or control packet, it immediately brings its receiver into promiscuous mode (GPSR Agent::tap), so as to overhear its immediate neighbour forwarding the packet [11]. The sending node verifies the different fields in the forwarded IP packet for requisite modifications through a sequence of integrity checks (GPSR Agent::verify packet integrity). If the integrity checks succeed, it confirms that the node has acted in a benevolent manner and so its direct trust counter is incremented. On the other hand, if the integrity check fails or the forwarding node does not transmit the packet at all, then its corresponding direct trust measure is decremented so that the node is treated as malicious node. The S-GPSR is explained by using flow chart which is illustrated through Figure 3.

## 4. SIMULATION RESULTS

The trust and mobility model is implemented in the existing GPSR protocol to obtain the S-GPSR protocol. The S-GPSR protocol is simulated using Network Simulator-2.32[12] to emulate selective forwarding and sinkhole attack in the mobile sensor network. The network animator output for 150 nodes with 10 malicious nodes indicated in red circle is shown in Figure 4. The performance parameters such as delivery ratio, delay and routing overhead are calculated for two different number of nodes (150 and 200) by varying the number of malicious nodes from 5 to 25 with various coverage areas such as 300×300 ($m^2$) and 500×500($m^2$). The parameters used in the simulation are listed in Table 1.

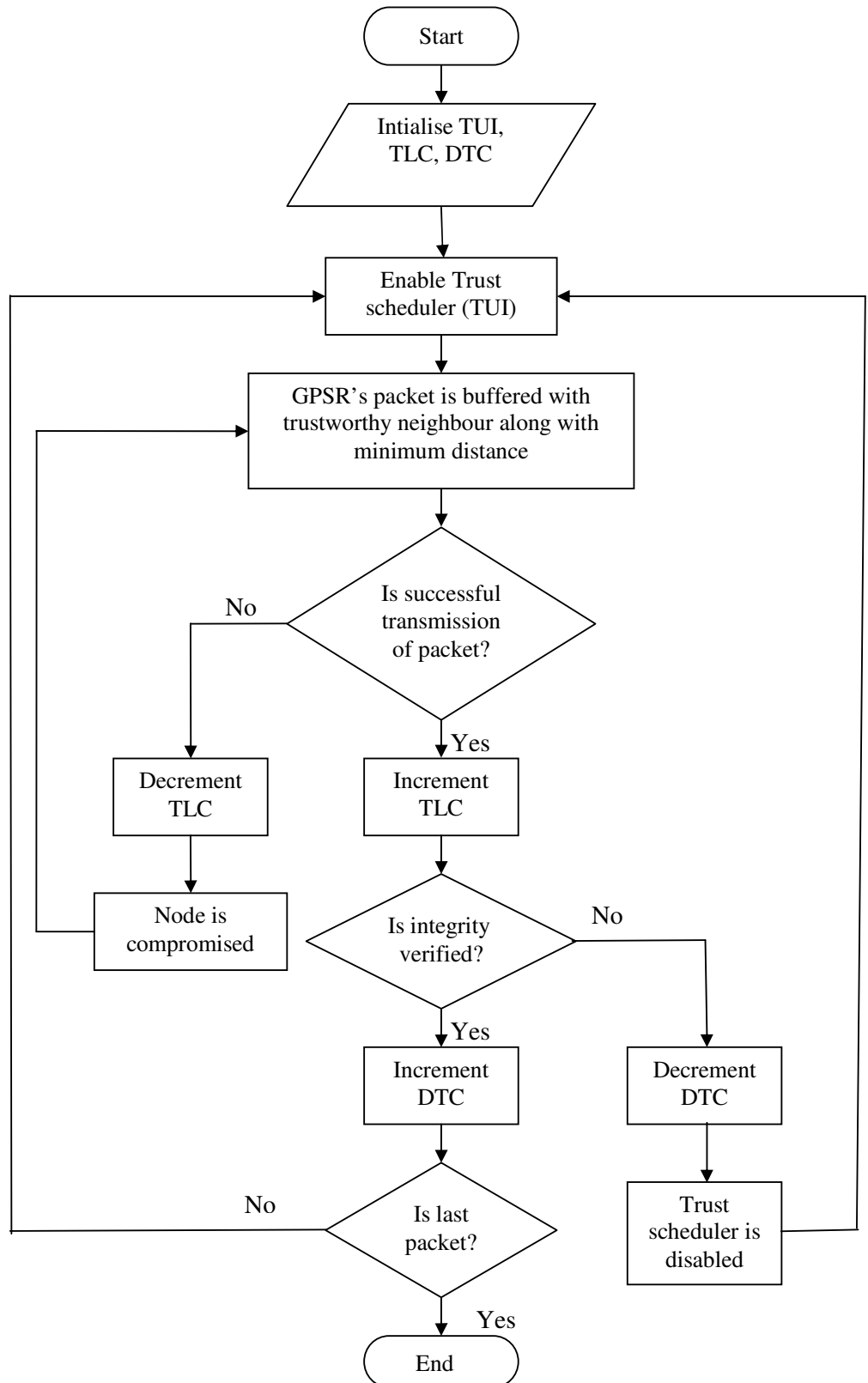

Figure 3. Flowchart of S-GPSR

Table.1 Simulation parameters

| Simulation Parameters | Values |
|---|---|
| Number of Nodes | 150 and 200 |
| Geographical area($m^2$) | 300×300, 500×500 |
| Packet Size(bytes) | 512 |
| Traffic Type | CBR |
| Number of malicious nodes | 5 to 25 |
| Mobility model | Random way point |
| Pause time(s) | 20 |
| Simulation time(s) | 100 |

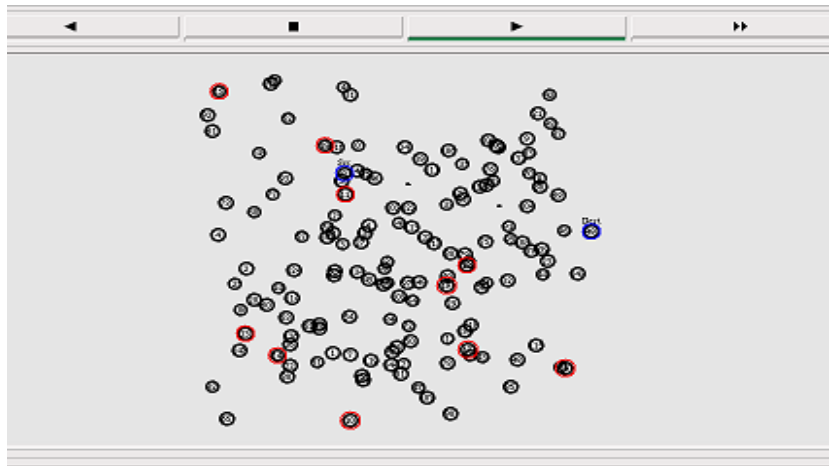

Figure 4.Network animator output for 150 nodes with 10 malicious nodes

### 4.1. Delivery Ratio

Delivery ratio of S-GPSR is higher than that of GPSR for 150 and 200 nodes with different coverage area of 300x300($m^2$) and 500x500 ($m^2$) which is shown in Figure 5, Figure 6, Figure 7 and Figure 8. The delivery ratio of S-GPSR is almost 98% up to 10 malicious nodes which is far better than GPSR which is illustrated by Figure 5. As the number of malicious nodes is increased further, the delivery ratio of S-GPSR is nearly 27% greater than that of the GPSR for 150 nodes. For 200 nodes, S-GPSR outperforms GPSR by providing delivery ratio of nearly 98% up to 15 malicious nodes.

On increasing the number of malicious nodes further S-GPSR provides a comparable increment of nearly 25% than that of GPSR which is shown in Figure 6. The fact is that shorter routes are preferred for transmitting the packets from source to destination in S-GPSR. Moreover, S-GPSR selects or deselects the neighbour node for routing process based on their trust levels to avoid the malicious node. Thus S-GPSR improves the forwarding rate by increasing the delivery ratio.

International Journal of Ad hoc, Sensor & Ubiquitous Computing( IJASUC ) Vol.1, No.2, June 2010.

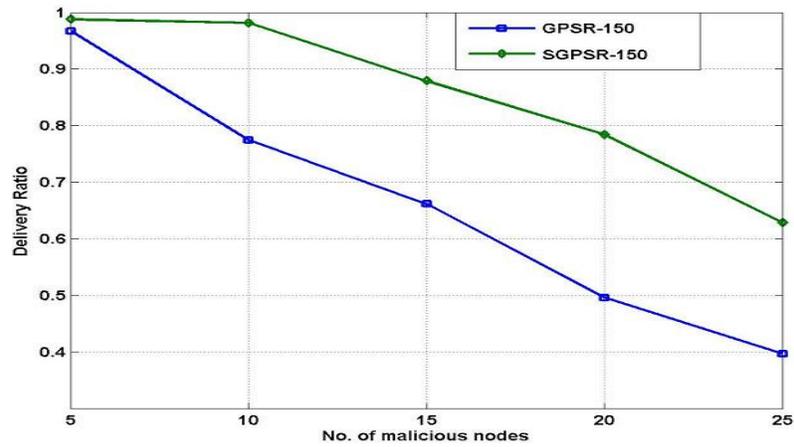

Figure 5. Delivery ratio with respect to no. of malicious nodes for 200 nodes with coverage area 300x300 m$^2$

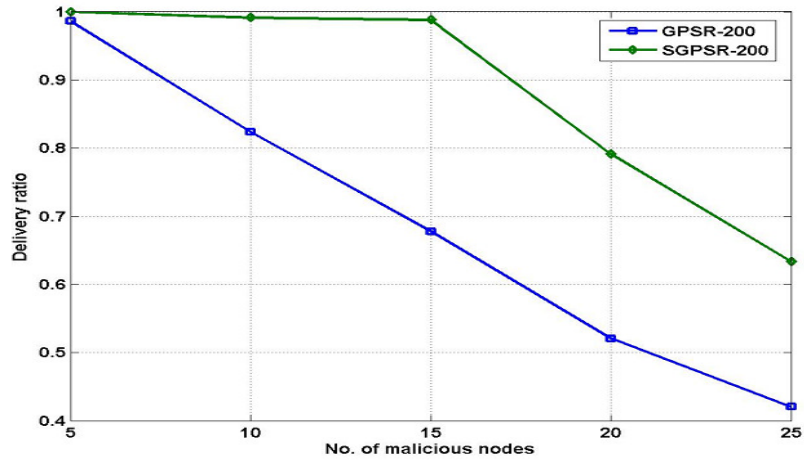

Figure 6. Delivery ratio with respect to no. of malicious nodes for 200 nodes with coverage area 300x300 m$^2$

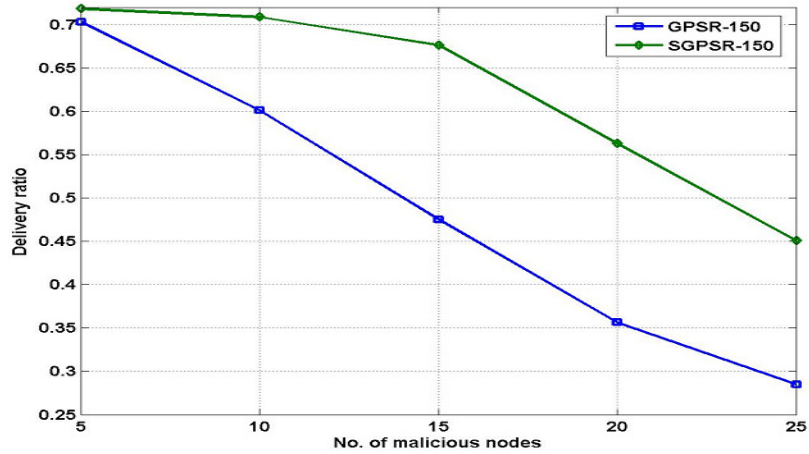

Figure 7. Delivery ratio with respect to no. of malicious nodes for 150 nodes with coverage area 500x500 m$^2$

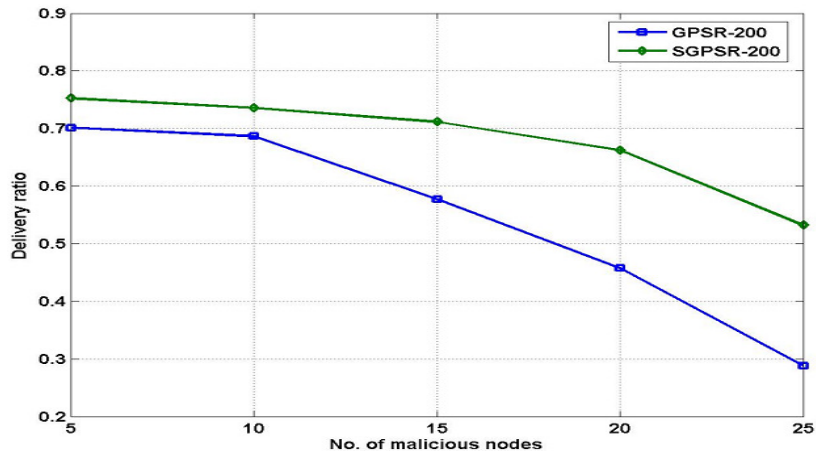

Figure 8. Delivery ratio with respect to no. of malicious nodes for 200 nodes with coverage area 500x500 m$^2$

## 4.2. Routing Overhead

It is the ratio between total numbers of control packets generated to total number of data packets received during simulation time. S-GPSR has an overall lower routing overhead compared to that of GPSR. This is illustrated through the results revealed in Figure 9, Figure 10, Figure 11 and Figure 12.

Though routing overhead of S-GPSR and GPSR increases for increased values of malicious nodes, S-GPSR achieves significant reduction in routing overhead of nearly 73% compared to that of GPSR. The reduced overhead is due to less number of control packets generated for each data packet in S-GPSR.

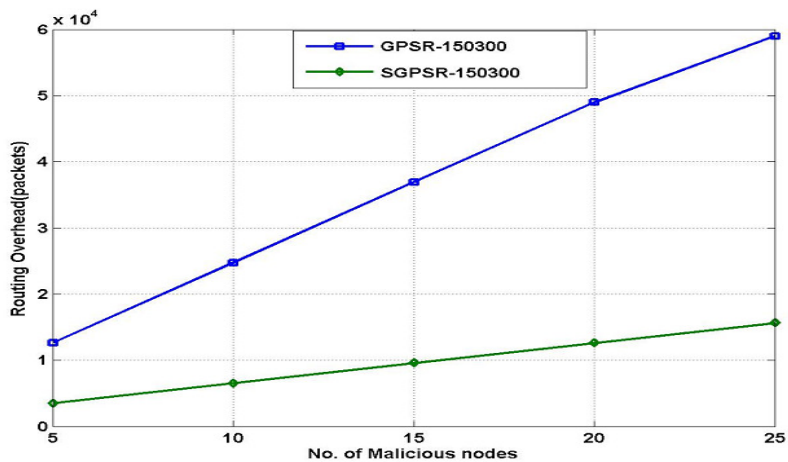

Figure 9. Routing overhead with respect to no. of malicious nodes for 150 nodes with coverage area 300x300 m$^2$

International Journal of Ad hoc, Sensor & Ubiquitous Computing( IJASUC ) Vol.1, No.2, June 2010.

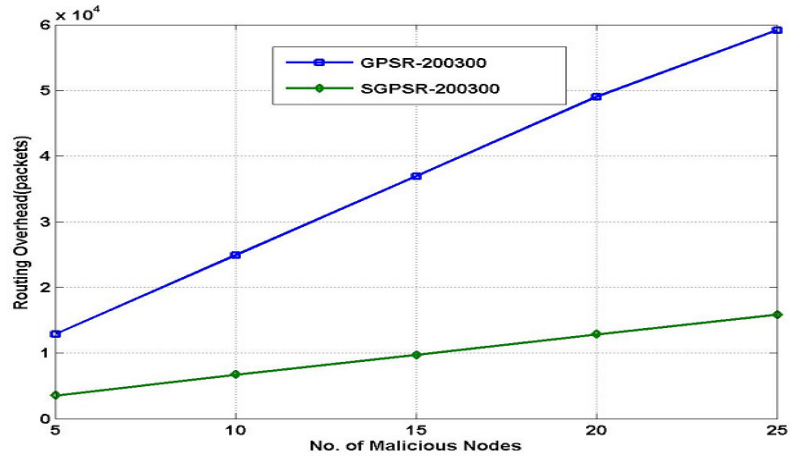

Figure 10. Routing overhead with respect to no. of malicious nodes for 200 nodes with coverage area 300x300 m$^2$

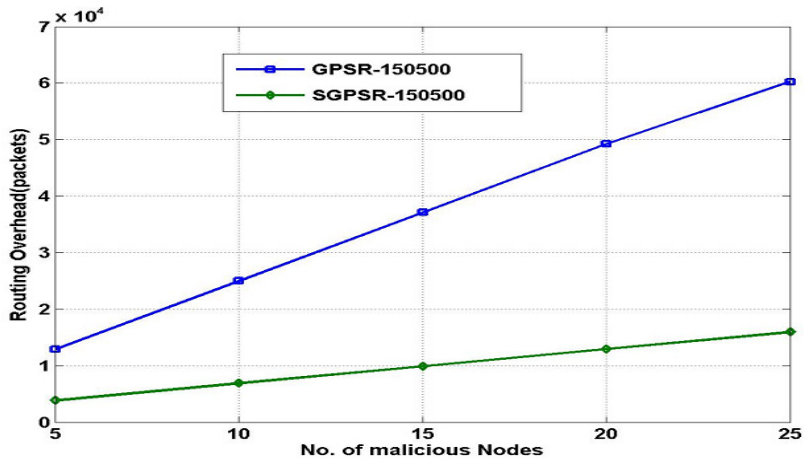

Figure 11. Routing overhead with respect to no. of malicious nodes for 150 nodes with coverage area 500x500 m$^2$

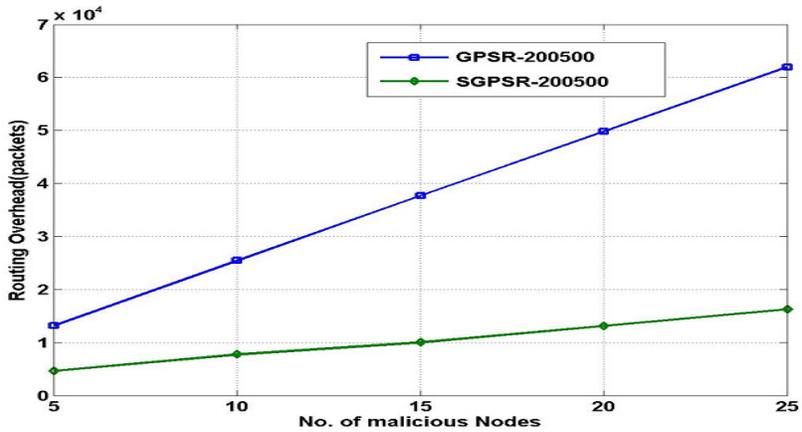

Figure 12. Routing overhead with respect to no. of malicious nodes for 200 nodes with coverage area 500x500 m$^2$

## 4.3. Delay

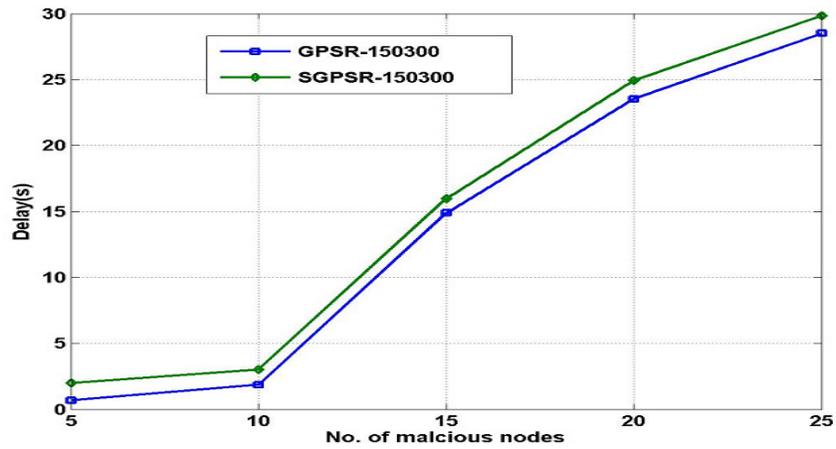

Figure 13. Delay with respect to no. of malicious nodes for 150 nodes with coverage area 300x300 m$^2$

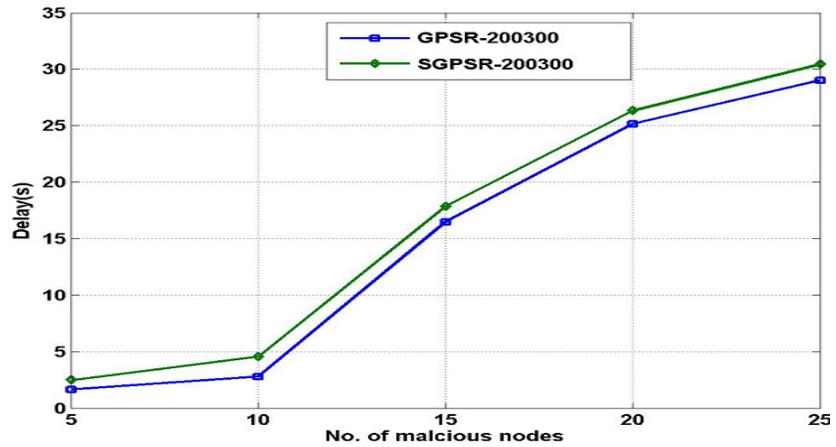

Figure 14. Delay with respect to no. of malicious nodes for 200 nodes with coverage area 300x300 m$^2$

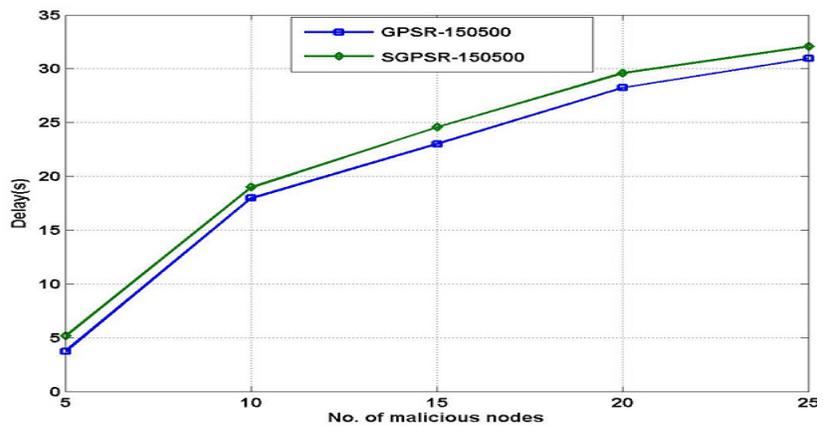

Figure 15. Delay Vs no. of malicious nodes for 150 nodes with coverage area 500x500 m$^2$

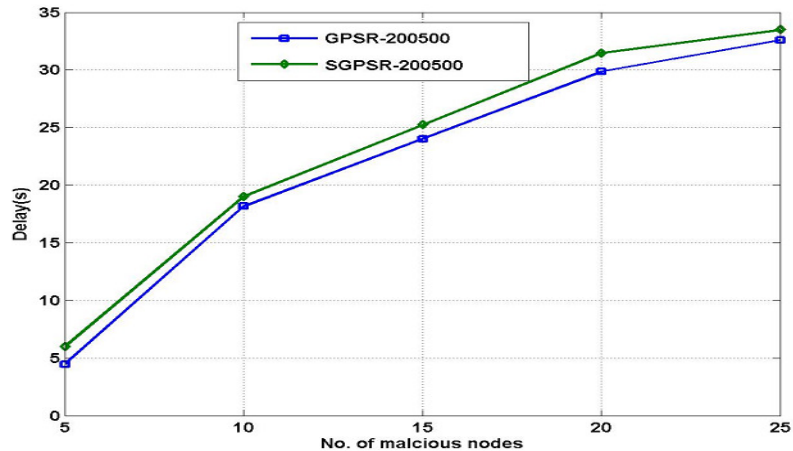

Figure 16. Delay Vs no. of malicious nodes for 200 nodes with coverage area 500x500 m$^2$

It is verified through simulation results shown in Figure 13, Figure 14, Figure 15 and Figure 16 that delay of S-GPSR protocol is higher than that of GPSR protocol. When numbers of malicious nodes are increased further, S-GPSR increases delay by 4% than that of GPSR protocol. S-GPSR selects intermediate nodes based upon their trusted path in addition to the minimal distance from the destination.

## 5. CONCLUSION

Secured greedy perimeter stateless routing protocol is implemented for mobile sensor network with different coverage area considering 150 and 200 number of nodes for simulation. It is compared with greedy perimeter stateless routing protocol for different number of malicious nodes. The results show that on the average, the routing overhead achieved using the S-GPSR protocol was 73% less than the standard GPSR protocol. Further more, an improvement of 25% in the delivery ratio have been achieved in the S-GPSR protocol. The improvement in the above mentioned network performance is mainly due to smaller trust values, shorter routing decisions and less number of control packets taken by the trust based model implemented in GPSR to get rid of the attackers.

**Authors**

**P. Samundiswary** received the B.Tech degree (1997) and M.Tech degree (2003) in Electronics and Communication Engineering from Pondicherry Engineering College, India. She is pursuing her Ph.D.  programme in the Dept. of Electronics and Communication Engineering, Pondicherry Engineering College affiliated to Pondicherry University, India. She is currently working as Assistant Professor in the Dept. of Electronics and Communication Engineering at Sri Manakula Vinayagar Engineering College affiliated to Pondicherry University, India. Her research interests include wireless communication and wireless sensor networks.

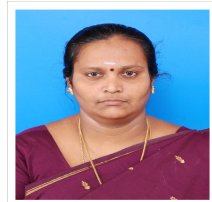

**D.Sathian** is pursuing his Bachelor of Technolgy in Computer Science and Engineering at Sri Manakular Engineering College, Pondicherry. His area of interest includes wireless networks, software Engineering.

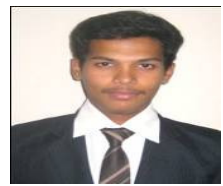

**P.Dananjayan** received Bachelor of Science from University of Madras in 1979, Bachelor of Technology in 1982 and Master of Engineering in 1984 from the Madras Institute of Technology, Chennai and Ph.D. degree from Anna University, Chennai in 1998. He is working as a Professor in the Dept. of ECE, Pondicherry Engineering College, India. He is also a Visiting Professor to AIT, Bangkok. He has more than 70 publications in National and International Journals. He has presented more than 150 papers in National and International conferences. He has produced 9 Ph.D. candidates and is currently guiding six Ph.D. students. His areas of interest include Spread spectrum Techniques and Wireless Communication

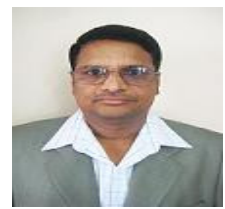